\documentclass[aps,prc,preprint,groupedaddress,showpacs,preprintnumbers,
amsmath,amssymb]{revtex4}

\usepackage{graphicx}
\usepackage{dcolumn}
\usepackage{bm}

\newcommand{\bra}{\langle}
\newcommand{\ket}{\rangle}
\newcommand{\sixj}[6]{
     \left\{ \begin{array}{ccc}
              #1 & #2 & #3 \\
              #4 & #5 & #6 
            \end{array}  \right\} } 
\newcommand{\ninej}[9]{
     \left\{ \begin{array}{ccc}
              #1 & #2 & #3 \\
              #4 & #5 & #6 \\
              #7 & #8 & #9
            \end{array}  \right\} } 
\begin{document}



\title{ 
Test of J-matrix inverse scattering potentials on electromagnetic reactions 
of few-nucleon systems
}



\author{Nir Barnea,}  
\affiliation{The Racah Institute of Physics, The Hebrew University,  
91904 Jerusalem, Israel}
\author{Winfried Leidemann and Giuseppina Orlandini}
\affiliation{ 
Department of Physics, The George Washington University, Washington DC 20052, USA\\
and Istituto Nazionale di Fisica Nucleare, Gruppo Collegato di Trento}
\altaffiliation{On leave of absence from Department of Physics, University of Trento, I-38050 Povo (Trento) Italy}

\date{\today}

\begin{abstract}
The J-matrix inverse scattering nucleon-nucleon 
potentials (JISP), describing both two-nucleon  data and bound and resonant 
states of light nuclei to high accuracy, are tested on the total photoabsorption 
cross sections of $^2$H, $^3$H and $^{3,4}$He. 
The calculations in the three- and four-body systems are carried out via the Lorentz 
integral transform method  and the  hyperspherical harmonics (HH)
technique. To this end the HH formalism has been adapted to accommodate non-local
potentials. The cross sections calculated with the JISP  are compared 
to those obtained with more traditional realistic interactions, 
which include two- and three-nucleon forces.
While the results of the two kinds of potential models do not differ
significantly at lower energies, beyond the resonance peak they show 
fairly large discrepancies, which increase with the nuclear mass.
We argue that these discrepancies may be due to a probably incorrect
long range behavior of the JISP, since the one pion exchange 
is not manifestly implemented there.

\end{abstract}



\maketitle


\section{\label{Introduction} Introduction}


In the last few years the possibility to perform very accurate calculations in
few-nucleon systems has stimulated a new attitude in studying the nature and 
form of the nuclear force. Differently from what was common in the past, nowadays 
potential models are tested not only on the increasingly accurate nucleon-nucleon 
(NN) data and triton binding energy, but also on bound and resonant states 
of more complex systems.  

The modern  debate on nuclear forces focuses on potential models that can 
schematically be grouped into three different categories. 
Category A includes the most traditional potentials. Their two-body parts 
are either based on meson exchange models ~\cite{BonnRA,Bonn,Nijmegen}, 
or largely phenomenological ~\cite{AV14,AV18}. However, all of them include 
the essential one pion exchange term at long range and reproduce the NN 
scattering data with very high accuracy. However, when tested on $A\geq 3$  
nuclei, they show non negligible underbinding. Three-body forces have been  
introduced to fit the  A=3 binding energy, and more 
recently also to describe energies of bound and low-lying resonant states
of p-shell nuclei ~\cite{TM,UVIII,UIX,IL2}. They require very intense 
computational efforts when used in may-body systems.

Category B includes the effective field theory (EFT) potentials \cite{EFT}, which have
a stronger connection to Quantum Chromo-Dynamics (QCD). They imply consistent 
two-, three- or more-body terms. The number of these terms and of the associated 
parameters, as well as  the accuracy in reproducing  data, increase significantly 
with the order of the expansions. The common feature of 
potentials belonging to categories A and B is the presence of three-body
operators. These require very intense computational efforts when used in many-body 
systems. 

Category C includes potentials that, while describing NN scattering data with 
high accuracy make use of the remaining off-shell freedom in order to reproduce 
few-body ground state energies and low-lying resonances, avoiding the necessity 
of three-body operators ~\cite{Doleschall,JISP6,JISPlong,JISP16,UCOM}. 

Electromagnetic reactions can provide good constraints of the off-shell behavior 
of the potential. The potentials of category A have been tested at large on quite 
a number of observables in the photo- and electro-disintegration of the two-body system. 
The comparison with data is very good, at least within the 
non relativistic regime (see e.g. ~\cite{ScA, AhLT}). 
As to category B potentials, analogous systematic studies  are still missing.
Investigations on electromagnetic reactions have been performed 
also in the three-body systems (see e.g. \cite{RLNOI, Benchmark, Golak, epelbaum})
for potentials falling into category A and B. 

Due to its large binding 
energy the four-body system represents a very good testing ground for off-shell 
versus many-body properties of the nuclear force. However, only recently an ab initio 
calculation 
of the total photodisintegration cross section of $^4$He has been performed \cite{PRLNOI}
with a category A  potential of local nature (AV18 and UIX) \cite{AV18, UIX}. 
Even if the lack of precise experimental 
data limits the validity of such kind of calculations as test of the off-shell 
behavior of the potentials, it is interesting to calculate the same observable also with a
potential of category C. In particular, one can ask whether  potential models 
that reproduce the binding energy of the alpha particle in two different ways,
either by including  three-body operators, or by a proper choice of the off-shell
behavior of the two-body potential, predict the same results for the
electromagnetic observables.

Among the potentials of category C, of particular interest are the potentials 
of Refs.~\cite{JISP6,JISPlong,JISP16}, which are highly non-local and given in terms of 
matrix elements 
constructed by means of the J-matrix version of inverse scattering theory.
Due to the lack of the three-body operator and the way they are constructed, they
lead to rapid convergence, when used in many-body ab initio 
calculations based on complete basis expansions, as was shown in \cite{JISPlong}
within the no core shell model approach \cite{NCSM}. 

The aim of this work is twofold. On the one hand we want to extend 
the calculation of Ref.~\cite{PRLNOI} to non-local potentials. On the other hand 
we would like to compare the results on the total 
photodisintegration of $^2$H, $^3$H and $^{3,4}$He below pion threshold, 
obtained with two versions of the JISP, to those given by typical category A potentials.

The choice of this particular electromagnetic observable is that in this case the validity of 
Siegert's theorem  allows to take largely into account, in an implicit way, 
the two- (or more-) body current contributions. 
Because of the particular form of the JISP  (in matrix and not in 
operator form) it would be otherwise very difficult, if not impossible, to 
construct them in a consistent way. On the other hand, just 
these currents can help to get an idea on the importance of the underlying 
degrees of freedom that are "mocked" by the off-shell part of the potential. 

The photodisintegration cross sections are calculated here using the 
Lorentz Integral Transform (LIT) method ~\cite{ELO94} which allows the 
treatment of the continuum 
states dynamics by means of a bound state technique. This method is briefly 
summarized in Sec. II.
We use the symmetrized HH expansion
\cite{symHH1,symHH2,reversedJacobi,symHH_akiva}  for the three- and four-body
calculations, while standard numerical methods are used to solve the differential 
equations in the two-body case. 

The use of the HH formalism is natural in 
configuration space and for local potentials. 
Since the JISP are non-local and given in terms of two-body
matrix elements between harmonic oscillator (HO) states, the many-body HH formalism 
has to be adapted to this case. Sec. III contains the description of how this is
achieved.
In Sec. IV we present the results for the total photodisintegration of 
A=2,3,4 nuclei and compare them to those obtained by category A potentials.
Conclusions are drawn in Sec. V. 


\section{\label{LIT} The total photoabsorption cross section with the LIT method}

The total photoabsorption cross section  is
given by
\begin{equation}
\sigma_\gamma(\omega)= 4 \pi^2 \alpha \omega R(\omega)\,,
\end{equation}
where $\alpha$ is the fine structure constant, $\omega$ represents the energy 
transferred by the photon and $R(\omega)$ is the response function defined as 
\begin{equation}\label{response}
   R(\omega)=\sum\!\!\!\!~\!\!\!\!\!\!\!\!\int _f\,\,
             |\bra f|\Theta| 0 \ket |^2 \delta(E_f -\omega-E_0) \, .
\end{equation} 
Here $|0\rangle$ and
$E_0$ are the nuclear ground state wave function and energy, $|f\rangle$ 
and $E_f$ denote eigenstates and eigenvalues of 
the nuclear Hamiltonian $H$, and $\Theta$ is the operator relevant to this
reaction (it is assumed that recoil effects are negligible). 
In the low-energy region considered here one
can rely on Siegert's theorem and use for $\Theta$ the unretarded dipole operator
\begin{equation}\label{operator}
\Theta=\frac{1}{2}\sum_i^A z_i \tau_i^3\,.
\end{equation}

The calculation of $R(\omega)$ seems to require the knowledge 
of the continuum states $|n\rangle$. However, this difficulty 
can be avoided using the LIT method. 
This method has been described extensively in several publications 
\cite{ELO94, ELO99}. Here we only summarize the three steps which 
are needed for the calculation of $R(\omega)$. 

{\it Step 1.} 
The equation  
\begin{equation} \label{LITeq}
(H- E_0-\omega_0 + i \Gamma)| \tilde \Psi \rangle = \Theta |0\rangle
\end{equation}
has to be solved for many $\omega_0$ and a fixed $\Gamma$.
This is a Schr\"odinger-like equation with a source. It can be shown easily 
that the solution $|\tilde\Psi \rangle$ is localized.
Thus  one only needs a bound state technique to calculate it. One generally
adopts the same bound state technique as for the solution of the ground state, which
is an input for Eq.~(\ref{LITeq}). We use expansions on the HH basis, as 
explained in Sec. III.

The values of the parameters $\omega_0$ and $\Gamma$ are chosen in relation 
to the physical problem. In fact, as it becomes clear in {\it Step 2}, the 
value of $\Gamma$ is a kind of ``energy resolution'' for the response function
and the values of $\omega_0$ scan the region
of interest. In our case we are interested in the resonance region and up to 
pion threshold. Therefore
we solve Eq.~(\ref{LITeq}) with $\Gamma=$ 10 and 20 MeV 
and for a few hundred of $\omega_0$ values chosen in the interval from 
$-10$ MeV to $200$ MeV.

{\it Step 2.} After solving Eq. (\ref{LITeq}) the overlap
$\langle\tilde\Psi|\tilde\Psi\rangle$ is
calculated. Of course this overlap
depends on $\omega_0$ and $\Gamma$. A theorem for integral transforms 
based on the closure property 
of the Hamiltonian eigenstates \cite{EFROS85} ensures that this dependence 
can be expressed as \cite{ELO94}
\begin{equation}\label{deflit}
 {\rm L} (\omega_0,\Gamma)=\langle\tilde\Psi|\tilde\Psi\rangle= \int R(\omega)\, 
          {\mathcal L}(\omega,\omega_0,\Gamma)\, 
d\omega\,,
\end{equation} 
where ${\mathcal L}$ is the Lorentzian function centered at
$\omega_0$ and with $\Gamma$ as a width:
\begin{equation}
{\mathcal L}(\omega,\omega_0,\Gamma)=
  \frac{1}{(\omega -\omega_0)^2 + \Gamma^2}\,.
\end{equation}
Therefore by solving Eq.~(\ref{LITeq}) one can easily obtain the LIT
 of the response function. 

{\it Step 3.} The transform (\ref{deflit})
is inverted (see e.g. Ref.~\cite{ELO99,ALRS05})
in order to obtain the response function  and therefore the
cross section. Of course the inversion result has to be independent of $\Gamma$ and
show a high degree of stability. 

\section{\label{sec:app}  Accommodation of non-local NN potential matrix elements
in the HH formalism}
In this section we describe the implementation of non-local potentials in the 
HH formalism. Our method is alternative to that of Ref.~\cite{Pisa_pspace}
and more convenient for the structure of our codes.


\subsection{The $A$-body  basis states }


In the HH formalism the antisymmetric $A$-body configuration-spin-isospin basis functions 
with total angular momentum $J_A,J^z_A$ and isospin $T_A,T^z_A$
are given by (notations as in  Ref.~\cite{HHnotation}) 

\begin{eqnarray} \label{HH_A}
| n_A {K}_A J_A J^z_A T_A T^z_A \Gamma_A \alpha_A \beta_A \rangle  & = &
\cr & & \hspace{-35mm}
      |{n_A}
\rangle\sum_{Y_{A-1}}
          \frac{\Lambda_{\Gamma_{A},Y_{A-1}}}{\sqrt{| \Gamma_{A}|}} \,
          \left[ | K_A L_A M_A \Gamma_A Y_{A-1} \alpha_A 
\rangle
            | S_A S^z_A T_A T^z_A \, \widetilde{\Gamma}_{A},\widetilde{Y}_{A-1}
                  \, \beta_A 
\rangle 
          \right]^{J_A}_{J^z_A} \,,
\end{eqnarray}
where
\begin{equation}\label{8}
 \langle \Omega_A | K_A L_A M_A \Gamma_A Y_{A-1} \alpha_A 
\rangle
      \equiv 
 {\cal Y}^{[A]}_{K_A L_A M_A \Gamma_A Y_{A-1} \alpha_A}(\Omega_A)   
\end{equation}
are the HH functions with hyperspherical angular momentum $K_A$, 
and orbital angular momentum quantum numbers $L_A, M_A$ that belong
to well defined irreducible representations (irreps) 
$\Gamma_{1} \in \Gamma_2 \ldots \in \Gamma_A $ of the permutation 
group-subgroup chain  
${\cal S}_1 \subset {\cal S}_2 \ldots \subset {\cal S}_A $,
denoted by the Yamanouchi symbol 
$[ \Gamma_A, Y_{A-1} ] \equiv [ \Gamma_A,\Gamma_{A-1},\ldots,\Gamma_1 ]$.
The dimension of the irrep $\Gamma_{m}$ is denoted by $| \Gamma_{m}|$
and $\Lambda_{\Gamma_{A},Y_{A-1}}$ is a phase factor \cite{Akiva88}.
Similarly, the functions 
\begin{equation}\label{9}
   \langle s^z_1..s^z_A, t^z_1..t^z_A 
|  S_A S^z_A T_A T^z_A \, \widetilde{\Gamma}_{A},\widetilde{Y}_{A-1} 
                                                      \beta_A\rangle 
   \equiv 
   \chi^{[A]}_{S_A S^z_A T_A T^z_A \, \widetilde{\Gamma}_{A},\widetilde{Y}_{A-1}
                  \, \beta_A}(s^z_1..s^z_A, t^z_1..t^z_A)
\end{equation}
are the symmetrized spin-isospin basis functions.
In the definitions (\ref{8}) and (\ref{9}) the quantum numbers $\alpha_A,\beta_A$ 
are used to remove the degeneracy 
of the HH and spin-isospin states, respectively.
The function
\begin{equation}
  \langle \rho | n_A 
\rangle = R_{n_A}^{[A]}(\rho)
\end{equation}
is the $A$-body hyperradial basis function,
\begin{equation}
   R_{n_A}^{[A]}(\rho) = \sqrt{\frac{n_A!}{(n_A+\alpha_L)!}}
   b^{-3(A-1)/2}\left(\frac{\rho}{b}\right)^{(\alpha_L-3A+4)/2}
   L_{n_A}^{\alpha_L}(\rho/b)\exp[-\rho/(2 b)]\,,
\end{equation}
where $L_n^{a}(x)$ are the associated Laguerre polynomials. The  $A$-body
hyperradial 
basis functions depend on the range parameter $b$
and the Laguerre parameter $\alpha_L$.

.


For the sake of brevity in the following the state of Eq.~(\ref{HH_A}) will be 
denoted by $|n_A, [K_A]\rangle$, with 
$[K_A]\equiv {K}_A J_A J^z_A T_A T^z_A \Gamma_A \alpha_A \beta_A $.

\subsection{The interaction and the HH $(A-2,2)$ basis }


The representation of a non-local two-body interaction becomes very
simple if the basis for the $A$-body Hilbert space is chosen as an outer 
product of a 2-particle  and a $(A-2)$-particle states.
In such a representation the interaction has the following form
\begin{equation}\label{pot}
v^{[2]}(ij)=\sum_{c^{}_{2} c'_{2} C_{A-2}} 
        |c_{2} C_{A-2} 
\rangle 
        v^{[2]}_{c^{}_{2} c'_{2}}\langle c'_{2} C_{A-2} | \,,
\end{equation}
where by $c_{2}$ and $c'_{2}$ we denote the two-body states of the interacting  
particles $ij$ and by $C_{A-2}$ we indicate a state that includes 
the $(A-2)$-body residual system. 
For a fermion system, these states should be taken as  
antisymmetric two-body and $(A-2)$-body states, respectively.

We choose to label the particles of the pair as particle $A$ and $A-1$ and to 
indicate their relative coordinate by $\vec{\eta} \equiv\vec{\eta}_{A-1}=
\sqrt{ \frac{1}{2}}(\vec r_A - \vec r_{A-1})$. 
Thus $c_{2}$ depends on the Jacobi coordinate $\vec{\eta}_{A-1}$ and $ C_{A-2}$ 
on the remaining $(A-2)$  reversed order Jacobi coordinates 
$\vec{\eta}_{A-2},\vec{\eta}_{A-3} ...\vec{\eta}_1$ ~\cite{reversedJacobi}.
Notice that in this way $C_{A-2}$ has information also on the relative 
orientation of the 2- and the $(A-2)$-body subsystems.


After having transformed $\{\vec{\eta}_{A-2}, \vec{\eta}_{A-3}\,  ... \,\vec{\eta}_1\}$  
and $\vec{\eta}$
to hyperspherical and spherical coordinates, respectively, one can express  
$C_{A-2}$ ($c_{2}$)
on the hyperspherical (spherical) basis. In analogy to Eq.~(\ref{HH_A})
one has the antisymmetric $(A-2)$-body configuration-spin-isospin states
\begin{eqnarray} \label{HH_{A-2}}
 &|&C_{A-2} 
  \rangle    =  
    | n_{A-2}
\rangle \sum_{Y_{A-3}}
    \frac{\Lambda_{\Gamma_{{A-2}},Y_{A-3}}}{\sqrt{| \Gamma_{A-2}|}} \,
    \cr & &
    \times\left[ | K_{A-2} L_{A-2} M_{A-2} \Gamma_{A-2} Y_{A-3} \alpha_{A-2} 
\rangle
           | S_{A-2} S^z_{A-2} T_{A-2} T^z_{A-2} 
           \, \widetilde{\Gamma}_{A-2},\widetilde{Y}_{A-3} \beta_{A-2} 
\rangle 
    \right]^{J_{A-2}}_{J^z_{A-2}} \,,
\end{eqnarray}
and the antisymmetric two-body states (the sum ${\ell_2+S_2+T_2}$ is odd)
\begin{equation}\label{c2}
|c_2 \rangle = | n_2\rangle \left[|{\ell_2 m_2}\rangle 
|S_2 S^z_2 T_2 T_2^z\rangle\right]^{J_2}_{J_2^z}\,.
\end{equation}

The notations in the previous equations are similar to those in Eqs. (\ref{HH_A}) 
and ({8}). The difference is in the subscripts $A-2$ and $2$, indicating that the 
quantum numbers refer to the $(A-2)$ system and to the pair, respectively.
Therefore the corresponding HH, spin-isospin and hyperradial functions are 
$ {\cal Y}^{[A-2]}_{K_{A-2} L_{A-2} M_{A-2} \Gamma_{A-2} Y_{A-3} 
\alpha_{A-2}}(\Omega_{A-2}) $ 
and
$ Y_{l_2 m_2}(\Omega_{2}=\hat \eta) $, 
$ \chi^{[A-2]}_{S^{}_{A-2} S^z_{A-2} T^{}_{A-2} T^z_{A-2} \widetilde{\Gamma}_{A-2}
\widetilde{Y}_{A-3} \beta_{A-2}}(s^z_1..s^z_{A-2}, t^z_1..t^z_{A-2}) $ and
$ \chi^{[2]}_{S_{2} S^z_{2} T_{2} T^z_{2}}(s^z_1 s^z_{2}, t^z_1 t^z_{2}) $, 
$ R_{n_{A-2}}^{[A-2]}(\rho_{A-2}) $ and 
$ R_{n_2 l_2}^{[2]}(\eta) $, respectively. Moreover, in the
following $ |C_{A-2}\rangle $ will be denoted by  
$ |n_{A-2},[K_{A-2}]\rangle $ and
$ |c_2 \rangle $ by 
$ |n_2,[K_2]\rangle $.



In order to calculate the matrix elements of the potential between
the $A$-body  states of Eq.~(\ref{HH_A}), 
it is also useful to introduce a particular HH basis
that reflects the division of the particle system into a pair of interacting
particles and an $(A-2)$ spectator. Such a basis is obtained by coupling
the 2-  and $(A-2)$-body 
spherical-spin-isospin states and hyperspherical-spin-isospin states, 
to yield an A-body basis function with quantum numbers
$[K_{A-2}],[K_2], K_A, J_A, J^z_A, T_A, T^z_A $, through the relation
\begin{eqnarray} \label{HH_2m2-0}
| [K_{A-2}]; [K_2]) K_A J_A J^z_A T_A T^z_A \ket & = & 
{\cal N}^{a, b}_{n} (\sin{\theta_{[A-2,2]}})^{\ell_2}
                     (\cos{\theta_{[A-2,2]}})^{K_{A-2}}
                     P^{(a,b)}_n(\cos{2\theta_{[A-2,2]}})\cr
 & & \times\left[ | [{K}_{A-2}] 
\rangle \; | [{K}_2] 
\rangle \right]^{J_A T_A}_{J^z_A T^z_A}\,,
\end{eqnarray}
%
%
%
where $\theta_{[A-2,2]}$ is defined by the following equations
\begin{eqnarray}\label{jacobirel}
  \eta &=& \rho\sin \theta_{[A-2,2]} \cr 
  \rho_{A-2}  &=& \rho \cos \theta_{[A-2,2]}\,.
\end{eqnarray}

Here $P^{(a,b)}_n$ are the Jacobi polynomials, with arguments
\begin{eqnarray}
   a & = & \ell_2+1/2                  \cr
   b & = & K_{A-2} + \frac{3A-8}{2}   \cr
   n & = & \frac{K_A-K_{A-2}-\ell_2}{2}   \;.
\end{eqnarray}
The numerical factor
\begin{equation}
{\cal N}^{a, b}_{n} = 
   \sqrt{\frac{2(2n+a+b) n! \Gamma(n+a+b+1)}
              {\Gamma(n+a+1)\Gamma(n+b+1)}
   }
\end{equation}
is a normalization constant.

For the sake of brevity we use the following notation:
$$ P_{K_A}^{K_{A-2},\ell_2}
\equiv
   {\cal N}^{a, b}_{n} (\sin{\theta_{[A-2,2]}})^{\ell_2}
                     (\cos{\theta_{[A-2,2]}})^{K_{A-2}}
                     P^{(a,b)}_n(\cos{2\theta_{[A-2,2]}})\,.
$$
Therefore the HH $(A-2,2)$ basis states
$| [K_{A-2}]; [K_2]) K_A J_A J^z_A T_A T^z_A \ket$
can now be written as
 \begin{equation} \label{HH_2m2}
| ([K_{A-2}]; [K_2]) K_A J_A J^z_A T_A T^z_A \ket =
P_{K_A}^{K_{A-2},\ell_2} \left[ | [{K}_{A-2}] 
\rangle \; | [{K}_2] 
\rangle \right]^{J_A T_A}_{J^z_A T^z_A}\,.
\end{equation}
These states, together with the $A$-body hyperradial basis states $|n_A\ket$,
form a complete orthonormal basis of our Hilbert space.


\subsection{The transformation between the 
$A$ and  $(A-2,2)$ basis }


The next step is the evaluation of the overlaps between 
the $A$-body functions, Eq.~(\ref{HH_A}), and the 
$(A-2)-$ and two-body functions, Eq.~(\ref{HH_{A-2}}) and (\ref{c2}).
To this end we use the
completeness of the HH $(A-2,2)$ basis, Eq.~(\ref{HH_2m2}), i.e. 
\begin{eqnarray}\label{overlap1}
\bra n_A,[K_A] \; |\; n_{A-2}, [K_{A-2}] ;\, n_2, [K_2]  \ket &=& \sum
\bra n_A,[K_A] \; |n_A, ([K_{A-2}]; [K_2]) K_A J_A J^z_A T_A T^z_A \ket
\cr && \hspace{-3cm}
\times\bra n_A ([K_{A-2}]; [K_2]) K_A J_A J^z_A T_A T^z_A \; 
       |\; n_{A-2}, [K_{A-2}] ;\, n_2, [K_2]  \ket
\;.
\end{eqnarray}
Let us start with the first matrix element on the right hand side of Eq.~(\ref{overlap1}).
The contributions 
of the hyperangular, spin and isospin 
matrix elements are evaluated with the help of $6j$ and $9j$ symbols, the 
hyperspherical coefficients of fractional parentage (CFPs)  and the 
spin-isospin CFPs \cite{BLO99}.
One has
\begin{eqnarray}
\lefteqn{
   \bra [K_A] | \left( [K_{A-2}] ; [K_2] \right) 
                K_A J_A J^z_A T_A T^z_A \ket = }
\cr & &
   \sqrt{(2J_{A-2}+1)(2J_2+1)(2S_A+1)(2L_A+1)}
   \ninej{S_2}{S_{A-2}}{S_A}{\ell_2}{L_{A-2}}{L_A}{J_2}{J_{A-2}}{J_A}
\cr & & \times
   \sum_{\Gamma_{A-1} }
   \Lambda_{\Gamma_{A},\Gamma_{A-1}}
   \Lambda_{\Gamma_{A-1},\Gamma_{A-2}}
   \sqrt{\frac{|\Gamma_{A-2}|}{|\Gamma_{A}|}}
\cr & & \times
   \langle K_A L_A Y_A \alpha_A | 
        \left( K_{A-2} L_{A-2} Y_{A-2} \alpha_{A-2} ; \ell_2  \right)
        K_A L_A 
\rangle 
\cr & & \times
   \langle S_A T_A \widetilde{Y}_A \beta_A | 
        \left( S_{A-2} T_{A-2} \widetilde{Y}_{A-2} \beta_{A-2} ; S_2 T_2  
        \right)
        S_A T_A 
\rangle\,,
\end{eqnarray}
where the  hyperspherical matrix elements are written as
\begin{eqnarray}
\langle K_A L_A Y_A \alpha_A | 
        \left( K_{A-2} L_{A-2} Y_{A-2} \alpha_{A-2} ; \ell_2 \right)
        K_A L_A 
\rangle  & = & 
\cr & & \hspace{-40mm} \sum_{ \alpha_{A-1}}
       \langle  K_A L_A \Gamma_{A-1} \alpha_{A-1} |
             K_A L_A \Gamma_{A} \alpha_{A}     
\rangle
\cr & & \hspace{-45mm} \times
       \langle  (K_{A-2} L_{A-2} \Gamma_{A-2} \alpha_{A-2}; \ell_2) K_A L_A |
             K_A L_A \Gamma_{A-1} \alpha_{A-1}     
\rangle
 \,
\end{eqnarray}
and, analogously, the  spin-isospin term is written as
\begin{eqnarray}
   \langle S_A T_A \widetilde{Y}_A \beta_A & | &
        \left( S_{A-2} T_{A-2} \widetilde{Y}_{A-2} \beta_{A-2} ; S_2 T_2 
        \right)
        S_A T_A 
\rangle =
\cr & & \hspace{-10mm}
   \sum_{S_{A-1} T_{A-1} \beta_{A-1}} \hspace{3mm} 
   \Pi \;TSCFP_A 
\cr & & \hspace{-10mm} \times
\langle S_{A-2}  S_{A-1}   S_{A} |
    (S_{A-2} ; S_{2}  )   S_{A} 
\rangle
\langle T_{A-2}  T_{A-1}   T_{A} |
    (T_{A-2} ; T_{2}   )   T_{A} 
\rangle\,,
\end{eqnarray}
with CFPs products
\begin{eqnarray}
\Pi \;TSCFP_A & = & 
    \langle  S_A S_{A-1} T_A T_{A-1} \widetilde\Gamma_{A-1} \beta_{A-1} |
          S_A T_A \widetilde\Gamma_{A} \beta_{A}     
\rangle
\cr & &
    \times \langle  S_{A-1} S_{A-2} T_{A-1} T_{A-2} \widetilde\Gamma_{A-2} \beta_{A-2} |
          S_{A-1} T_{A-1} \widetilde\Gamma_{A-1} \beta_{A-1}     
\rangle\,.
\end{eqnarray}
The spin-isospin matrix elements 
can  be easily evaluated using
angular momentum techniques:
\begin{equation} \label{T2m2}
    \langle T_{A-2}  T_{A-1}   T_{A} |
    (T_{A-2} ; T_{2}  )   T_{A} 
\rangle = 
        (-)^{T_{A}+T_{A-2}+1} 
        \sqrt{(2T_{A-1}+1)(2T_{2}+1)}
   \sixj{T_{A-2}}{t}{T_{A-1}}{t}{T_A}{T_{2}} \,.
\end{equation}
Here $t$ is equal to $\frac{1}{2}$ and  stands for the isospin of a single nucleon. 
The spin matrix element can be obtained by simply replacing 
in Eq.~(\ref{T2m2}) the isospin quantum numbers by corresponding spin 
quantum numbers.
The overlap between the interaction basis and the HH $(A-2,2)$ states
are given by 
\begin{eqnarray}\label{overlap2}
\bra n_A ([K_{A-2}]; [K_2]) K_A J_A J^z_A T_A T^z_A \; 
       |\; n_{A-2}, [K_{A-2}] ;\, n_2, [K_2]  \ket
&=& 
\cr && \hspace{-9cm} 
\bra J_{A-2} J_{A-2}^z J_2 J_2^z |  J_A J^z_A \ket
  \bra T_{A-2} T_{A-2}^z T_2 T_2^z |  T_A T^z_A \ket
  \Delta^{n_{A-2}\,n_{2}}_{n_A}(K_A,K_{A-2},{\ell_2}) 
\end{eqnarray}
where,
\begin{eqnarray}
\Delta^{n_{A-2}\,n_{2}}_{n_A}(K_A,K_{A-2},{\ell_2}) = 
\cr && \hspace{-4cm}
  \int  \eta^2 d\eta R^{[2]}_{n_{2} \ell_2}(\eta)
  \int \rho_{A-2}^{3A-7} d \rho_{A-2} \;
  R_{n_{A-2}}^{[A-2]}(\rho_{A-2}) R_{n_{A}}^{[A]}(\rho)
  P_{K_A}^{K_{A-2},\ell_2}(cos\theta_{[A-2,2]})\,.
\end{eqnarray}
This two dimensional integral can be easily evaluated using the
 Gauss-quadrature integration.


\subsection{The potential matrix elements }

Using the representation of the potential as in Eq.~(\ref{pot})
and the overlaps $ \langle n_A,[K_A] \; |\; c_2 \,C_{A-2}\rangle \; $
calculated above,
the $A$-body matrix elements of a scalar-isoscalar two-body operator can 
be written as
\begin{eqnarray}\label{v_aap}
& & \bra n_A [K_A] | v^{[2]}(A,A-1) \;| \;  n'_A [K'_A] \ket =   
\sum_{c^{}_{2} c'_{2} C_{A-2}}
\bra n_A [K_A] \; |c_{2} C_{A-2} \rangle 
    v^{[2]}_{c^{}_{2} c'_{2}}
\langle c'_{2} C_{A-2} | \;  n'_A [K'_A] \ket = \cr
 &=&\sum_{[K_2],[K'_2],[K_{A-2}]}
\sum_{[n_2],[n'_2],[n_{A-2}]}
\Delta^{n_{A-2}n_{2}}_{n_A}(K_A,K_{A-2},{\ell_2})   
\Delta^{n_{A-2}n'_{2}}_{n'_A}(K'_A,K_{A-2},{\ell'_2})  \cr
& \times &\bra [K_A] \;|\; (K_{A-2}];[K_2])K_A J_A J^z_A T_A T^z_A 
\ket 
  \bra [K'_A] \;|\; ([K_{A-2}];[K'_2])K'_A J_A J^z_A T_A T^z_A\ket
\cr & &
\;v^{[2]}_{n^{}_{2} [K^{}_2] , n'_{2} [K'_2]}(A,A-1)  \,,
\end{eqnarray}
%
where we have used the orthogonality of the Clebsch Gordon coefficients
\begin{equation}
\sum_{J_2^z J_{A-2}^z}
\bra J_{A-2} J_{A-2}^z J_2 J_2^z |  J_A J^z_A \ket
\bra J_{A-2} J_{A-2}^z J_2 J_2^z |  J'_A J^{\prime z}_A \ket
=\delta_{J_A J'_A}\delta_{J^z_A J^{\prime z}_A}
\end{equation}
and
\begin{equation}
\sum_{T_2^z T_{A-2}^z}
\bra T_{A-2} T_{A-2}^z T_2 T_2^z |  T_A T^z_A \ket
\bra T_{A-2} T_{A-2}^z T_2 T_2^z |  T'_A T^{\prime z}_A \ket
=\delta_{T_A T'_A}\delta_{T^z_A T^{\prime z}_A}\,.
\end{equation}

Finally, the actual potential matrix elements are calculated via
\begin{equation}
v^{[2]}_{n^{}_{2} [K^{}_2] , n'_{2} [K'_2]} = 
     \delta_{J^{}_2 J'_2}
     \delta_{J^{z}_2 J^{\prime z}_2}
     \delta_{T^{}_2 T'_2}
     \delta_{T^{z}_2 T^{\prime z}_2}
     \bra n^{}_{2} \ell^{}_2 S^{}_2 J^{}_2 J^z_2 T^{}_2 T^z_2
     |v^{[2]}|
          n'_{2} \ell'_2 S'_2 J^{}_2 J^z_2 T^{}_2 T^z_2 \ket\,.
\end{equation}

\subsection{Application to the J-matrix inverse scattering potentials}

In view of the use of the JISP  as given in ~\cite{JISP6,JISP16}, 
a convenient choice for the configuration space two-body basis 
functions are the HO states,
\begin{equation}\label{ho_nlm}
 \langle \vec r | n_2 \ell_2 m_2 
\rangle_{HO} = 
  R^{HO}_{n_2 \ell_2}(r)   Y_{\ell_2 m_2}(\hat r)\,,
\end{equation}
with the radial function
\begin{equation}\label{ho_r}
 R^{HO}_{n_2 \ell_2}(r) =
  \frac{(-1)^{n_2}}{r}
   \sqrt{\frac{2n!}{r_0\Gamma(n+l+3/2)}}\left(\frac{r}{r_0}\right)^{\ell_2+1}
   exp[-r^2/(2r_0^2)]L_{n_2}^{\ell_2+\frac{1}{2}}(r^2/r_0^2) \;.
\end{equation}
Here $r$ is the distance between the two particles i.e. 
$r\equiv |\vec r_1-\vec r_2|$,
$r_0$ is related to the oscillator strength $\Omega_{HO}$ and the reduced mass $\mu$
through the relation $r_0=\sqrt{\hbar/\mu\Omega_{HO}}$.
However, the basis states in Eq.~(\ref{ho_nlm}) do not possess the proper
normalization. In fact they are normalized according to 
\begin{equation}
  \int d\vec r |\langle \vec r | n l m 
\rangle_{HO} |^2 = 1\,,
\end{equation}
while in the canonical transformation from the single particle coordinates
to the centre of mass and reversed order Jacobi coordinates the 
appropriate normalization is
\begin{equation}
  \int d\vec \eta |\langle \vec \eta | n l m 
\rangle |^2 = 1\,,
\end{equation}
where 
$\vec\eta=\sqrt{ \frac{1}{2}}(\vec r_A - \vec r_{A-1}) 
         =\sqrt{ \frac{1}{2}} \vec r$.
Consequently a normalization factor of $ \sqrt[4]{8}$ has to be introduced.
Therefore the appropriate two-body basis functions are 
\begin{equation}\label{ho_nlmp}
 \langle \vec \eta | n_2 \ell_2 m_2 
\rangle = 
     R^{[2]}_{n_2 \ell_2}(\eta) Y_{\ell_2 m_2}(\hat \eta)
\equiv  \sqrt[4]{8}
     R^{HO}_{n_2 \ell_2}(r=\sqrt{2}\eta) 
      Y_{\ell_2 m_2}(\hat \eta)\,.
\end{equation}

\section{\label{Results} Results}

In the following we discuss the results for the unretarded total photoabsorption
cross section of the two-, three-, and four-nucleon systems obtained with the JISP 
and compare them to the results given by other nuclear force models. For the three- 
and four-body systems the cross sections are calculated with the LIT method, as outlined 
in Sec. II, and using HH expansions of ground state wave function and $\tilde\Psi$.
For the JISP these HH expansions have rather rapid convergences, 
though no effective interaction has been introduced in this case. We obtain convergent 
binding energies of the three- and four--body
systems with $K\leq 20$ (see table \ref{tb:be_ground_states}), while for potentials
like AV18 one needs $K$ larger than 100.
For the deuteron photoabsorption with the JISP  we also use the LIT method, 
but here combined with expansions on HO functions. For the other
NN potentials a conventional calculation with explicit NN final state wave functions
is carried out (for one of these potentials it was checked that we get the same
result also with the LIT method).    

Before discussing the total photoabsorption cross sections we first consider the 
longitudinal deuteron elastic form factor $F_L(q^2)$. 
In Fig.~1 we show results for the JISP6 \cite{JISP6}, 
AV14 \cite{AV14}, and BonnRA \cite{BonnRA} 
potential models (note that the proton and neutron electric form factors are set 
equal to 1 and 0, respectively). Up to a momentum transfer $q^2=2$ fm$^{-2}$
all models lead to rather similar results, while for higher $q^2$ the JISP6 
result is considerably lower than those of the other two potential models.
In comparison with experimental data one usually considers the elastic deuteron
form factors $A(q^2)$ and $B(q^2)$, where $A(q^2)$ contains besides the
longitudinal also a transverse contribution. However in the momentum range shown 
in Fig.~1 the transverse piece is very small (see, e.g. \cite{Galster}). For  
this momentum range it is known that the $A(q^2)$ of conventional potential 
models like AV14 or BonnRA agree well with experimental data (see e.g. \cite{AV14}).
Thus one may conclude that the JISP6 potential leads to a good description
of data only up to $q^2=2$ fm$^{-2}$. On the other hand this might be
sufficient for the calculation of electromagnetic low-momentum transfer reactions like
nuclear photoabsorption below pion threshold.

In Fig.~2 the unretarded deuteron total photoabsorption cross section is shown
for various potential models. In Fig. 2a one sees that the low-energy peaks
have slightly different heights (note that the cross section below about 
5 MeV is also affected by the here not considered M1 contribution). 
Beyond the peak one finds a good
agreement of all models up to about 10 MeV. At higher photon energies one has
very similar results for the AV14, AV18 \cite{AV18}, BonnRA models, 
while the JISP6 result shows
a somewhat higher cross section between 15 and 60 MeV (see Fig. 2b). Fig. 2c helps 
to understand the situation better. For the BonnRA potential
we illustrate the effect of the inclusion of retardation and of other 
multipoles (electric and magnetic ones up to multipole order $L$=4), where the
non-relativistic one-body current and the Siegert operator are considered (calculation 
corresponds to Arenh\"ovel's {\it normal} calculation for the deuteron 
photodisintegration, see e.g. \cite{SaA}). One notes
that the additional contributions increase the cross section only slightly.
In principle one has to include also other current contributions
(meson exchange, isobar, and relativistic currents), however, as shown
in Fig.~7.1.8 of Ref.~\cite{SaA} these effects are rather small. In the considered
energy range they lead to a reduction of the {\it normal} cross section
between 2\% and 3\%, thus bringing the total result very 
close to the unretarded total photoabsorption cross section.
In comparison to experimental data \cite{Ahe74,SkS74,Bob79,DPG82,BeI86} there is good agreement of 
the unretarded cross section with the BonnRA potential, while
with the JISP6 potential the data are somewhat overestimated .

Now we turn to the total photoabsorption cross section of $^3$H and $^4$He. 
In Fig.~3a we show the triton case with the
JISP6 and JISP16 potentials and for the nuclear force models AV14 \cite{AV14} 
and UVIII \cite{UVIII}; AV18 \cite{AV18} and UIX \cite{UIX}; BonnRA \cite{BonnRA} and 
Tucson-Melbourne (TM) \cite{TM}. Results for the AV14+UVIII and BonnRA+TM potentials 
are taken from Ref.~\cite{ELOT00}. The situation is similar to the deuteron case
of Fig.~2: small differences among the low-energy peak heights of all potential
models and beyond 20 MeV differences between the conventional nuclear force models 
and the JISP, which, however, are considerably more pronounced than for
the deuteron case.
Since beyond the peak there are no triton total photoabsorption cross section 
data available, the comparison with data is done for the $^3$He case (see
Fig.~3b), choosing one of the JISP (JISP6) and one conventional (AV18+UIX) potential model.
Unfortunately the data \cite{datielio3} have rather large error bars and do not lead to a 
clear picture. However, one can say that between 20 and 50 MeV there is a better agreement 
with the JISP6, while beyond 50 MeV the conventional potential models are favored.

In Fig.~4 we show the $^4$He total photoabsorption cross section for the JISP6 and JISP16
potential models, as well as for the AV18+UIX nuclear force. The calculation for AV18+UIX
is described in \cite{PRLNOI}. There it is discussed that the convergence of the HH
expansion is rather slow because of the three-nucleon force. As already
mentioned, for the JISP we find a much better convergence, though no effective interaction
is introduced. in Fig.~4 one notices that the trend observed
for deuteron and three-nucleon cases is confirmed. Beyond the peak the cross sections 
obtained with the JISP are considerably higher than that obtained using
the conventional nuclear force model AV18+UIX. This effect is more pronounced 
than for the three-nucleon case, which in turn
is already stronger than found for the deuteron case.
In addition, for $^4$He the JISP lead  
to somewhat lower peak heights and to a shift of the peak position by about 2 MeV 
towards higher energy, when compared to the AV18+UIX result.
It is evident that the latter  agrees much
better with the experimental data \cite{Ark}. Here we should mention
that the experimental situation is not yet settled (see also \cite{PRLNOI}),
particularly regarding the height of the low-energy peak (the shown data are the
only experimental total $^4$He $\sigma_\gamma(\omega)$ results that extend
to energies beyond the peak region).

\section{\label{Conclusions}Conclusions}

In this work we have presented results for the total photoabsorption cross section of
the two- three- and four-nucleon systems, obtained within the recently proposed J-matrix 
inverse scattering potential models. 

The calculation in the three- and four-body systems are performed via the LIT method  
and the  HH technique. To this aim the HH formalism has been adapted to accommodate 
non-local potentials. 

The comparison of the JISP results with those obtained
with a few more traditional realistic potentials, including two- and three-body forces, 
has shown that, while the latter give very similar results, JISP
display a rather different behavior, especially beyond the resonance peak. 
The differences increase with the nuclear mass. 

These results show that 
a "classical" electromagnetic observable as the total photonuclear cross section 
is able to emphasize the rather different off-shell properties of these two classes 
of potentials.
In particular, in this observable non-localities of the two-body potential and
three-body forces are not equivalent.
Considering that, different from the JISP,
 the traditional potentials have in common the long range one-pion exchange part, 
one may ascribe the discrepancies to this fact.
 
In the case of deuteron the comparison with data speaks in 
favor of the more traditional two-body potentials. 
For the three-body systems data are missing or 
have insufficient accuracy, making the comparison inconclusive. Regarding the 
four-body system the old data of Ref.~\cite{Ark} seem to favor the traditional 
potentials, as in the deuteron case. More recent data obtained in Lund \cite{Nilsson}, 
while measuring
only the $^4$He$(\gamma,n)$ cross section point in this direction as well, provided that
they are extrapolated to $\sigma_\gamma$ by means of the charge symmetry argument. 

More data of higher accuracy are very much needed to further clarify the 
issue of two- and three-body forces.

\section*{Acknowledgment}
We thank A.M. Shirokov for sending us the potential matrix elements
of the JISP models.
The work of N.B. was supported by the Israel Science Foundation
(grant no. 361/05).

\newpage


\begin{figure}
\resizebox*{15cm}{17cm}{\includegraphics[angle=0]{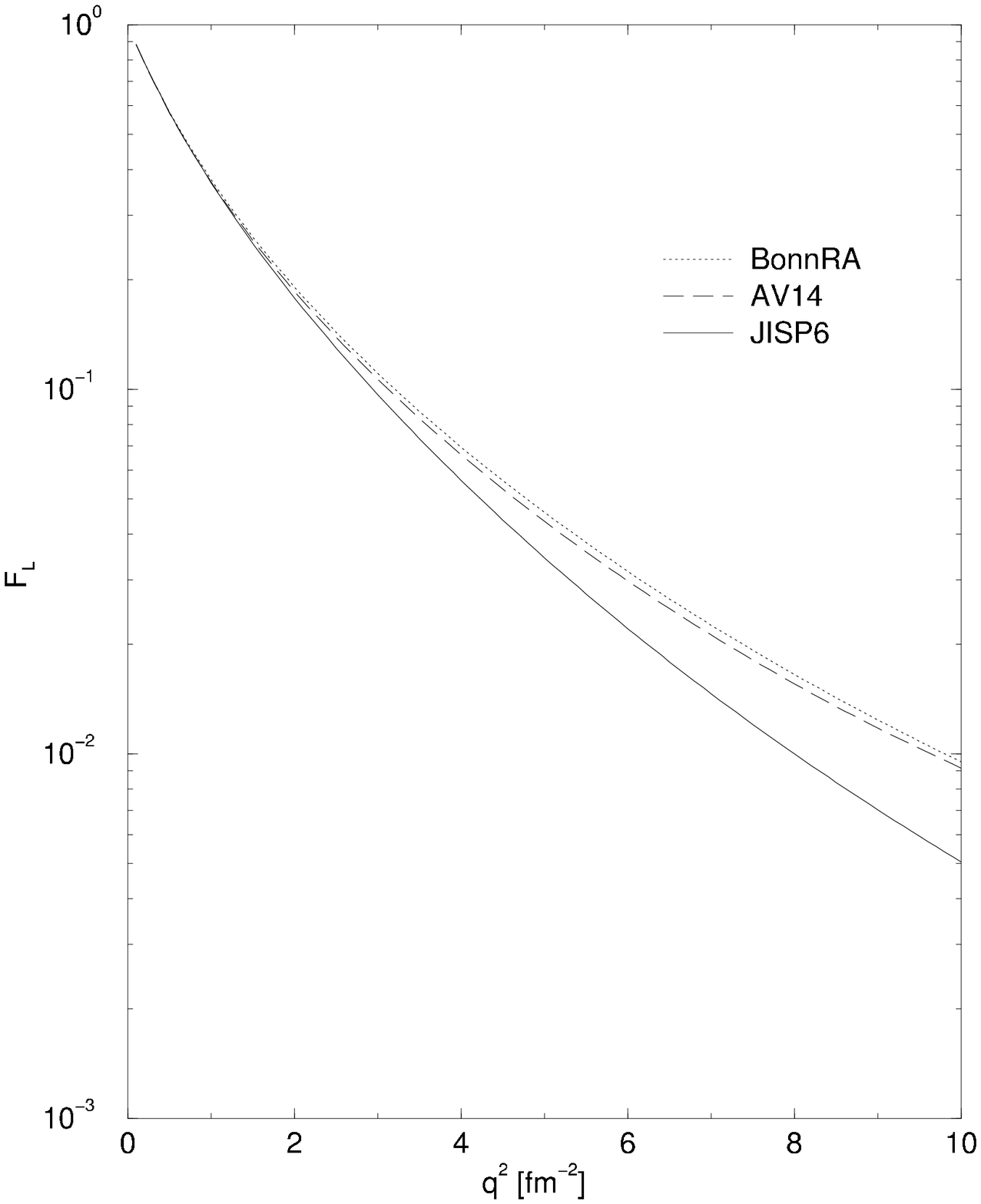}}
\caption{Deuteron elastic longitudinal form factor (see text)
for three different potential models.
} 
\label{figure1}
\end{figure}

\begin{figure}
\resizebox*{15cm}{17cm}{\includegraphics[angle=0]{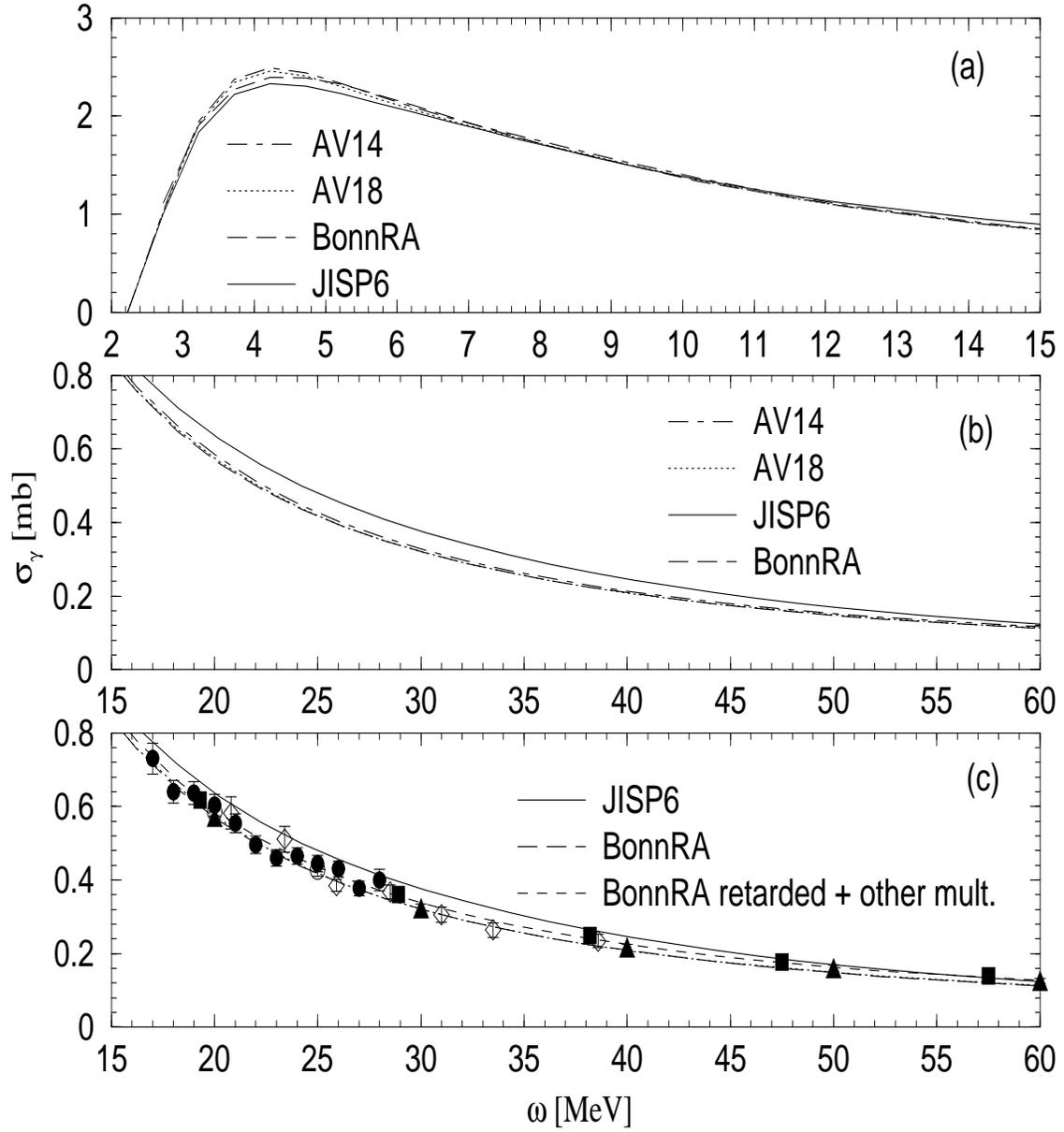}}
\caption{Deuteron total photoabsorption cross section at lower (a)
and higher (b) photon energies for four different potential models.
Comparison with data for two potentials (see text) is shown in (c): 
open circles~\cite{Ahe74}, 
full squares~\cite{BeI86}, open diamonds~\cite{Bob79}, 
full triangles~\cite{DPG82}, full circles~\cite{SkS74}.
} 
\label{figure2}
\end{figure}

\begin{figure}
\resizebox*{15cm}{17cm}{\includegraphics[angle=0]{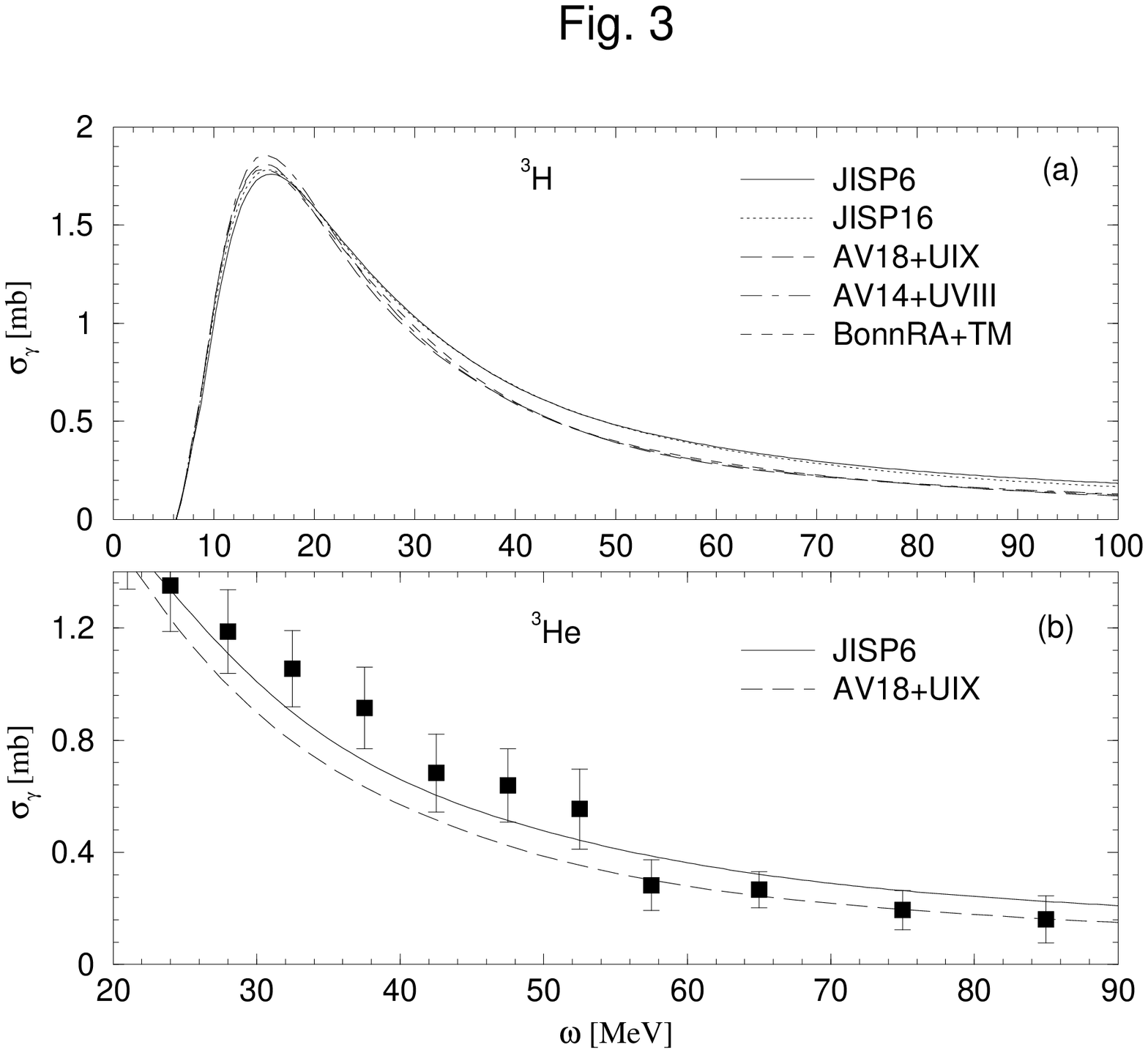}}
\caption{Three-body total photoabsorption cross sections: (a)  $^3$H results for five 
different potential models; (b): $^3$He results 
for two potential models, compared to data from Ref.~\cite{datielio3}.
} 
\label{figure3}
\end{figure}

\begin{figure}
\resizebox*{15cm}{17cm}{\includegraphics[angle=0]{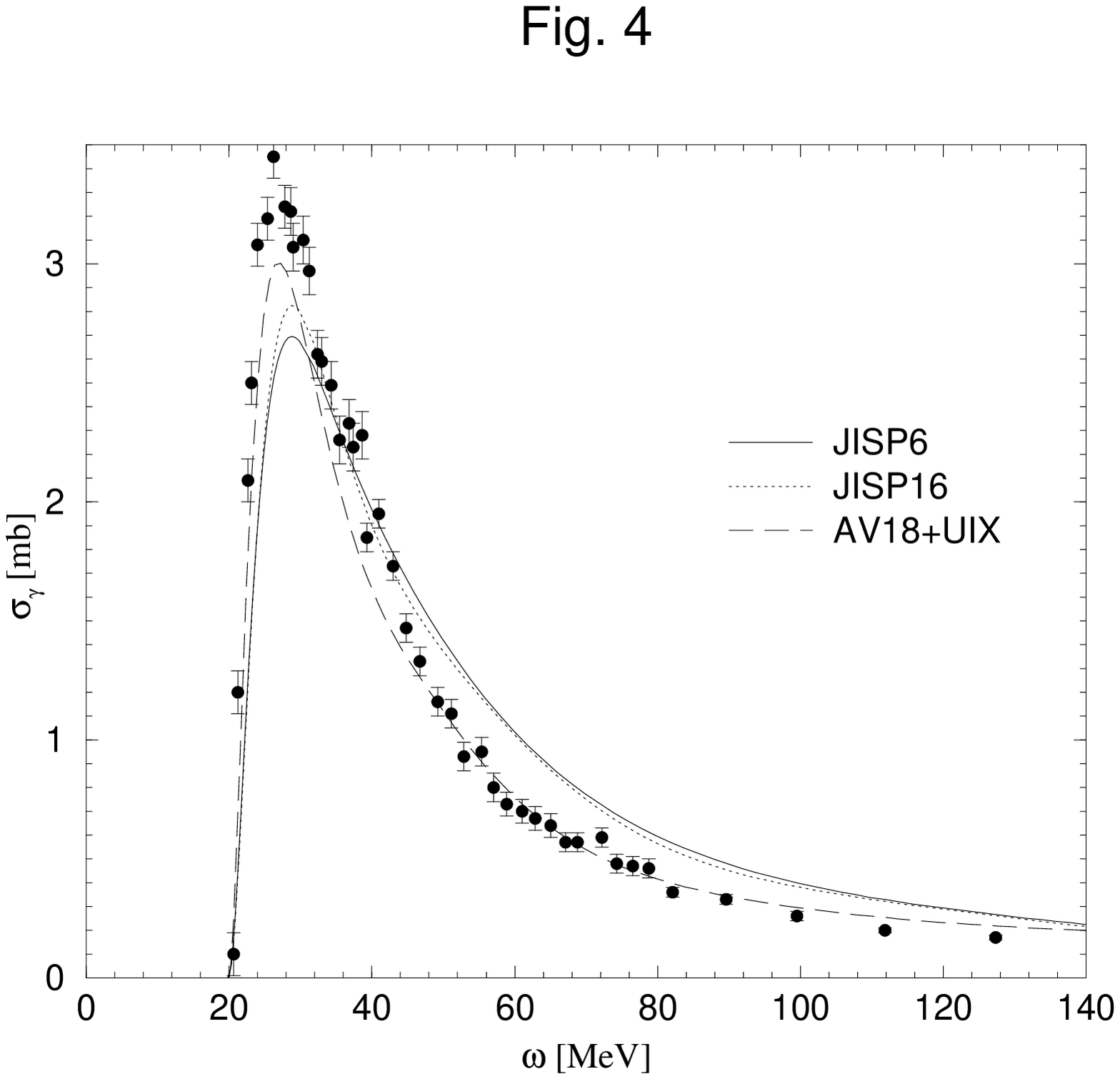}}
\caption{Total photoabsorption cross section of $^4$He for three different potential 
models. Experimental data from Ref.~\cite{Ark}.
} 
\label{figure4}
\end{figure}

\begin{table}
\caption{
\label{tb:be_ground_states}
Binding energies [MeV] for $A=3,4$ nuclei using the HH expansion 
with the JISP6 and JISP16 potentials. For comparison we also present the
results obtained in the NCSM approach.
}
\begin{center}
\begin{tabular}{ c | cc | cc}
                & \multicolumn{2}{c|}{\;\;\;\;\;\;JISP6 \;\;\;\;\;\;} 
                & \multicolumn{2}{c }{\;\;\;\;\;\;JISP16\;\;\;\;\;\;}\\
  Nucleus       &HH & NCSM \cite{JISP6} & HH & NCSM \cite{JISP16} \\
\hline \hline
  $^3$H   &  8.461(1)   & 8.461(5)    &  8.369(1) &  8.354  \\
  $^3$He  &  7.749(1)   & 7.751(3)    &  7.662(1) &  7.648  \\
  $^4$He  & 28.602(1)   & 28.611(41)  & 28.299(1) & 28.297  \\
\hline \hline
\end{tabular}
\end{center}
\end{table}
\end{document}